\documentclass[preprint, 12pt, a4paper,sort&compress]{elsarticle}

\usepackage{amssymb}
\usepackage{amsmath}
\usepackage{listings}
\usepackage{xcolor}
\usepackage{lineno,hyperref}	
\usepackage{float}
\restylefloat{table}

\usepackage{xparse}

\newcommand{\software}{\textsc{OptiMic}}
\newcommand{\highlight}{\color{blue}}
\newcommand{\Eref}[1]{Equation~(\ref{#1})}%
\newcommand{\sref}[1]{\emph{section}~\ref{#1}}%
\newcommand{\fref}[1]{\emph{fig}.~\ref{#1}}%

\NewDocumentCommand{\codeword}{v}{%
\texttt{\textcolor{blue}{#1}}%
}

\journal{SoftwareX}

\begin{document}

\begin{frontmatter}

\title{OptiMic: A tool to generate optimized polycrystalline microstructures for materials simulations}

\author[mimm]{P.~H.~Serrao}
\author[mimm,fzj,rwth]{S.~Sandfeld}
\author[mimm]{A.~Prakash\corref{cor}}
\ead{arun.prakash@imfd.tu-freiberg.de}
\cortext[cor]{Corresponding author}
\address[mimm]{Micromechanical Materials Modelling (MiMM), \\
Institute of Mechanics and Fluid Dynamcis, \\TU Bergakademie Freiberg, Lampadiusstra{\ss}e 4, 09599 Freiberg, Germany}
\address[fzj]{Institute for Advanced Simulation -- IAS-9: Materials Data Science and Informatics, \\Forschungszentrum Juelich GmbH, 52425 Juelich, Germany}
\address[rwth]{Chair of Materials Data Science and Materials Informatics,\\ Faculty 5 -- Georesources and Materials Engineering, \\RWTH Aachen University, 52056 Aachen, Germany}

\begin{abstract}

Polycrystal microstructures, with their distinct physical, chemical,  structural and topological entities, play an important role in determining the effective properties of materials. Particularly for computational studies, the well-known Voronoi tessellation technique is regularly used for obtaining microstructures. Standard Voronoi tessellations, however, exhibit statistics that are generally far removed  from those in real microstructures. Nevertheless, such tessellations can be optimized to obtain certain key features and statistics seen in real microstructures. In this work, we develop the open-source software package \software{} that enables the generation of optimized microstructures for both finite element as well as atomistic simulations. \software{} allows for both monodispersive grains as well as irregular grains obtained currently via Voronoi tessellations.  These initial microstructures can then be optimized to reflect desired statistical features. A key feature of the tool is that it gives the user extensive control on the optimization process via customizable cost functions. The software currently performs tessellations with the Voronoi method and can be easily extended to include other methods like grain-growth, phase-field etc.

\end{abstract}


\end{frontmatter}


\section*{Code metadata}
\label{code_metadata}

\begin{table}[H]
\begin{tabular}{|l|p{6.5cm}|p{6.5cm}|}
\hline
\textbf{Nr.} & \textbf{Code metadata description} & \textbf{} \\
\hline
C1 & Current code version & 2.0 \\
\hline
C2 & Permanent link to code/repository used for this code version & {OptiMic code repository \cite{optimicRepo}} \\
\hline
C3 & Code Ocean compute capsule & $NA$\\
\hline
C4 & Legal Code License   & GNU GPLv3 \\
\hline
C5 & Code versioning system used & git \\
\hline
C6 & Software code languages, tools, and services used & Python, Gmsh, nanoSCULPT, Ovito \\
\hline
C7 & Compilation requirements, operating environments \& dependencies & Numpy, Scipy, Matplotlib, Numba, Pytest, Numpy-quaternion, Tess, Click, gmsh-sdk, Yappi \\
\hline
C8 & If available Link to developer documentation/manual & OptiMic- Wiki \cite{optimicWiki} \\
\hline
C9 & Support email for questions & arun.prakash@imfd.tu-freiberg.de\\
\hline
\end{tabular}
\caption{Code metadata}
\label{} 
\end{table}


\section{Introduction}

Computer simulations have now become an integral part of studies on material behavior. Simulations are increasingly being used over the entire length scale to investigate and explain various phenomena \cite{raabe1998,raabe2004}. Particularly in studies dealing with the deformation behavior, multiscale simulation techniques are being regularly used to describe, for example, the thermo-mechanical response of a body subjected to various complex initial and boundary conditions \cite{guo2007multiscale}. Notwithstanding the underlying model used to describe the physics of the material as authentically as possible, the predictive power of such simulations depends on two further factors \cite{prakash2017}: i) the reproduction of the real-world boundary conditions in the simulated model, and ii) the inclusion of all relevant statistical and topological information in the underlying microstructure so as to accurately describe the effective material behavior.

The microstructure that we refer to is, at the most fundamental level, comprised of individual cells called grains. Each grain is characterized by a distinct orientation, size and shape \cite{janssens2010}. The microstructure is made up of further topological entities of varying dimensions: zero-dimensional (0-D) entities called vertex or quadruple points (0-D) are formed at the contact point of four or more grains, one-dimensional (1-D) entities called triple junctions (TJs) or lines (1-D) are formed at the intersection of three or more grains, and two-dimensional (2-D) entities called grain boundaries (GBs) are the interfaces between two grains \cite{xu2010,li2011}. Further entities include textural characteristics like orientation of individual grains, mis-(dis)orientation and orientation relationship between grains. Both structural and textural characteristics can significantly influence GB properties, topology and network, and thus affect the overall anisotropic response of the polycrystal.

It hence follows that inclusion of microstructure in simulations refers not only to the packing of cells, but also to the inclusion of relevant statistical information on further topological entities. Four kinds of models are generally used to describe the microstructure \cite{gross2002,prakash2016nanosculpt}: a) Monodispersive grain size models (e.g., \cite{stukowski2010, frydrych2019,amodeo2016}, where all grains have the same shape like hexagons or cubes; b) Voronoi based tessellations (e.g., \cite{carson2017,brandl2011}), c) microstructures obtained from physics based models of grain-growth, like e.g. phase-field, vertex grain growth, Potts monte carlo etc. \cite{janssens2010,syha2009,mason2015,prakash2017NC}, and d) digitization of realistic microstructures via e.g. serial sectioning \cite{alkemper2001,uchic2011}.

Of these, the usage of Voronoi based microstructures is almost ubiquitous in both finite element as well as atomistic simulations, due to the relative ease in generating such structures. Although the grains obtained by Voronoi tessellations look similar to those seen in experimental micrographs, closer quantitative analysis shows that many topological and statistical properties are quite removed from those observed in real microstructures \cite{gross2002,prakash2017NC}. Nonetheless, some microstructural statistics can be tailored by using optimization algorithms so as to obtain optimal initial structures for further simulations. This is done by minimizing a cost function, which penalizes any deviation of the desired statistic from the target distribution.
 
In this work, we develop a software tool called \software{} that generates optimized microstructures for both finite element as well as atomistic simulations. The tool generates both mono-dispersive as well as Voronoi tessellations, and allows for the optimization of desired statistics so as to reflect probability distributions as specified by the user. The software computes a variety of statistical information than can be used for the optimization process. \software{} is a part of growing ecosystem of tools, e.g. \textsc{Neper} \cite{neper}, \textsc{Kanapy} \cite{kanapy}, \textsc{dream3d} \cite{dream3d}, \textsc{MicroStructPy} \cite{microstructpy} that enable the generation of optimal microstructures mainly for finite element simulations.  A key feature of \software{} is that it gives the user extensive control on the optimization via customizable cost functions, which can in turn be used to treat specific microstructural features. Additionally, the modular approach used in the development allows an easy extension of the code for further tessellation methods like vertex grain growth, phase field etc.


\section{Software description}
Generation of a microstructure with \software{} involves the following steps:

\begin{itemize}
\item Generate seeds/grain centers in a user defined simulation box
\item Tessellate simulation box via the Voronoi method using seed positions generated previously
\item Construct cost function using desired metric
\item Optimize microstructure by moving the position of seeds
\item Write out seed coordinates, structural and textural statistics, and other relevant information
\item Generate finite element mesh
\item Output surface mesh in VTK \cite{schroder2000vtk} and OBJ \cite{murray2005wavefront,bourkeOBJ} file format for generation of atomistic structures and for further processing
\end{itemize}

The initial step in the generation of the desired microstructure is to tessellate the simulation box with a pre-defined number of seeds or grain centers, which are either placed at regular intervals or at random positions. In the current version of \software{}, the well-known Voronoi method is used for the tessellation. The Voronoi tessellation generates grain domains $\mathcal{T}$ that enclose all points with closest distances to the corresponding grain center. Formally, this can be represented as:
\begin{equation} \label{eq:voronoi_formula}
\mathcal{T}_p = \{ \boldsymbol{\mathnormal{x}} \in \mathcal{D} \, | \parallel \boldsymbol{\mathnormal{x}} - S_p \parallel < \parallel \boldsymbol{\mathnormal{x}} - S_q \parallel \}, \hspace{1em} \textrm{with} \hspace{0.5em} p, q = 1  \ldots  N_g \hspace{0.5em} \textrm{and} \hspace{0.5em} p \neq q, 
\end{equation}
where $\boldsymbol{\mathnormal{x}} $ denotes the coordinates of any point in the $n$-dimensional simulation box $\mathcal{D}$, $N_g$ is the total number of grains, and $S$ denotes the set of coordinates of grain centers. With grain seeds placed at regular intervals, this method results in mono-dispersive grains; with random seeds, one obtains the standard Poisson-Voronoi tessellation.

The so-obtained tessellation is then optimized in a subsequent step. In general, the objective of the optimization process is to achieve a target distribution of a particular metric. The cost function $\mathcal{C}$ for optimization is defined simply as the sum of squared differences between the current and target distributions, $Y$ and $\hat{Y}$, respectively:
\begin{equation} \label{eq:CostFunc}
\mathcal{C} = \frac{1}{m}\sum_{i = 1}^m (Y_i - \hat{Y_i})^2,
\end{equation} 
where $m$ is the number of points (or bins) chosen to discretize the distribution. In the current version of \software{}, the metric chosen for optimization can be distributions of any of the following: i) grain sizes, ii) number of neighbors, iii) grain boundary areas, iv) triple junction lengths, v) angles at a triple junction, vi) disorientation angles, vii) type of GB (in terms of sigma values), and viii) maximum Schmid factors.  

\Eref{eq:CostFunc} is the cost function used by default in \software{}; the user may, however, also define a custom cost function. To this end, in addition to the above mentioned metrics in terms of distributions, the distance between grain centers is also made available.
An example of a custom cost function is shown as part of application examples in \sref{sec:2Ducf}.

The cost function is now minimized using optimization algorithms, which essentially move the grain centers to more favorable positions so as to reflect the desired distribution of the chosen metric. Both gradient-based and gradient-free algorithms can be used for this purpose. The final optimized microstructure is then meshed and/or filled with atoms for further simulations.

\subsection{Software details}
\software{} is written in Python and uses the following libraries to accomplish various tasks: 

\begin{itemize}
\item \textbf{Voro++} \cite{rycroft2009voro++} for generation of Voronoi tessellations. The Python package \textbf{Tess} \cite{tess} is used to interface with Voro++.
\item \textbf{scipy.optimize} \cite{scipy} for optimization of the microstructure
\item \textbf{Gmsh} \cite{geuzaine2009gmsh}, together with its Python API, for generation of tetrahedral and hexahedral meshes for finite element simulations. 
\item \textbf{nano\textsc{sculpt}} \cite{prakash2016nanosculpt} for generation of the atomistic sample.
\end{itemize}

The generated microstructure is periodic in all directions. \software{} can generate microstructures in both 2D and 3D. The precise execution instructions, including all options of the program, can be found on the wiki page \cite{optimicWiki} of \software{}. Although all optimization algorithms in \textbf{scipy.optimize} \cite{scipy} are available to the user, currently \software{} has only been tested with the \textsc{Cobyla} \cite{cobyla} and \textsc{SLSQP} \cite{kraft1989slsqp} algorithms.

Once the microstructure has been optimized, the volume defining individual grains is written out as a surface mesh in the VTK \cite{schroder2000vtk} and OBJ \cite{murray2005wavefront} file formats; these files are then used for generating atomistic configurations using \emph{nano\textsc{sculpt}} \cite{prakash2016nanosculpt}. Furthermore, all relevant statistical information is written out for further processing by the user. The seed coordinates and orientations of individual grains are written out every iteration, thus enabling a restart of the optimization process, if needed.

\section{Application examples}

The following examples were generated on a system with AMD Ryzen 5 2500U (4 physical cores, 8 virtual cores) with Radeon Vega Mobile Gfx processor with 8 Gigabytes (GB) of RAM. The CPU architecture is equipped with L1d, L1i, L2 and L3 cache memory of 32 Kilobytes (KB), 64 KB, 512 KB and 4096 KB respectively. All examples were generated with the simplex-based \textsc{Cobyla} optimization algorithm \cite{cobyla}.

\subsection{3D microstructure with log-normal grain size distribution}
In this example, we generate a 3D microstructure with 500 grains in a simulation box of size \mbox{15 x 15 x 15} units. The aim of the optimization process in this example is to ensure a (close to) log-normal grain size distribution; herein, the maximum function evaluations were set to 2000. Each grain has a face-centered cubic (fcc) crystal structure and is given a random orientation. The following command is used for execution: 

\begin{lstlisting}[language=bash]
$> <@\highlight{./optimic main --size}@> 15. 15. 15. <@\highlight{--dimension}@> 3 
   <@\highlight{--number\_seed}@> 500 <@\highlight{--target}@> user_grain_size_distribution.txt 
   <@\highlight{--characteristic}@> 0 <@\highlight{--material}@> example_1_3D 
   <@\highlight{--stress\_direction}@> 1 0 0 <@\highlight{--seed\_spacing}@> random_3d 
   <@\highlight{--rand\_seed}@> 1 <@\highlight{--optimization\_method}@> COBYLA <@\highlight{--max\_iter}@> 2000 
   <@\highlight{--number\_bins}@> 20 <@\highlight{--mesh}@> hex
\end{lstlisting}

\begin{figure}[htbp!]
\centering
\includegraphics[width=0.95\textwidth]{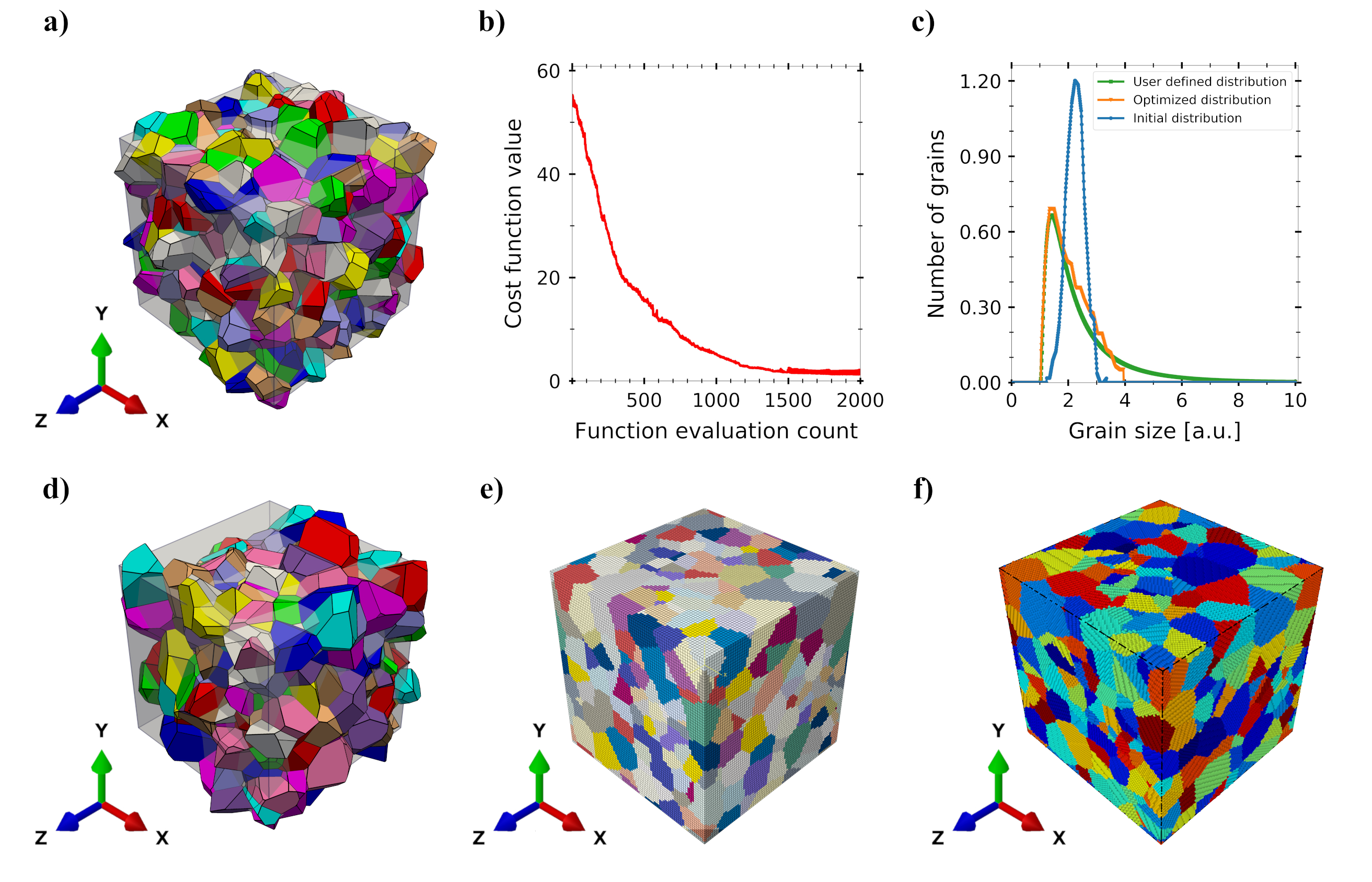} 
\caption{Generation of a 3D microstructure with 500 grains in a simulation box of size \mbox{15 x 15 x 15} units  using \software{} and optimized to a target log-normal grain-size distribution: a) Initial Voronoi tessellation with randomly placed seeds; b) Evolution of cost function during the optimization process; c) Initial, target and optimal grain size distribution; d) Voronoi tessellation after optimization of grain size distribution; e) Finite element mesh; f) Atomistic configuration. The transparent cube in a) and d) shows the limits of the simulation box. Individual grains are colored with different colors in a, d-f.} 
\label{fig:3Dmicro}
\end{figure}

By default with random seeds, the Voronoi method generates a tessellation with Poisson-Voronoi grain size distribution (cf. \fref{fig:3Dmicro}a,b). The \textsc{Cobyla} algorithm \cite{cobyla} is able to successfully reduce the cost function and result in a microstructure with log-normal grain size distribution (see \fref{fig:3Dmicro}b,c). The final microstructure along with the meshed and atomistic configurations are shown in \fref{fig:3Dmicro}d,e,f. The optimization process is shown as a movie in supplementary movie S1.

\begin{figure}[htbp!]
\centering
\includegraphics[width=0.95\textwidth]{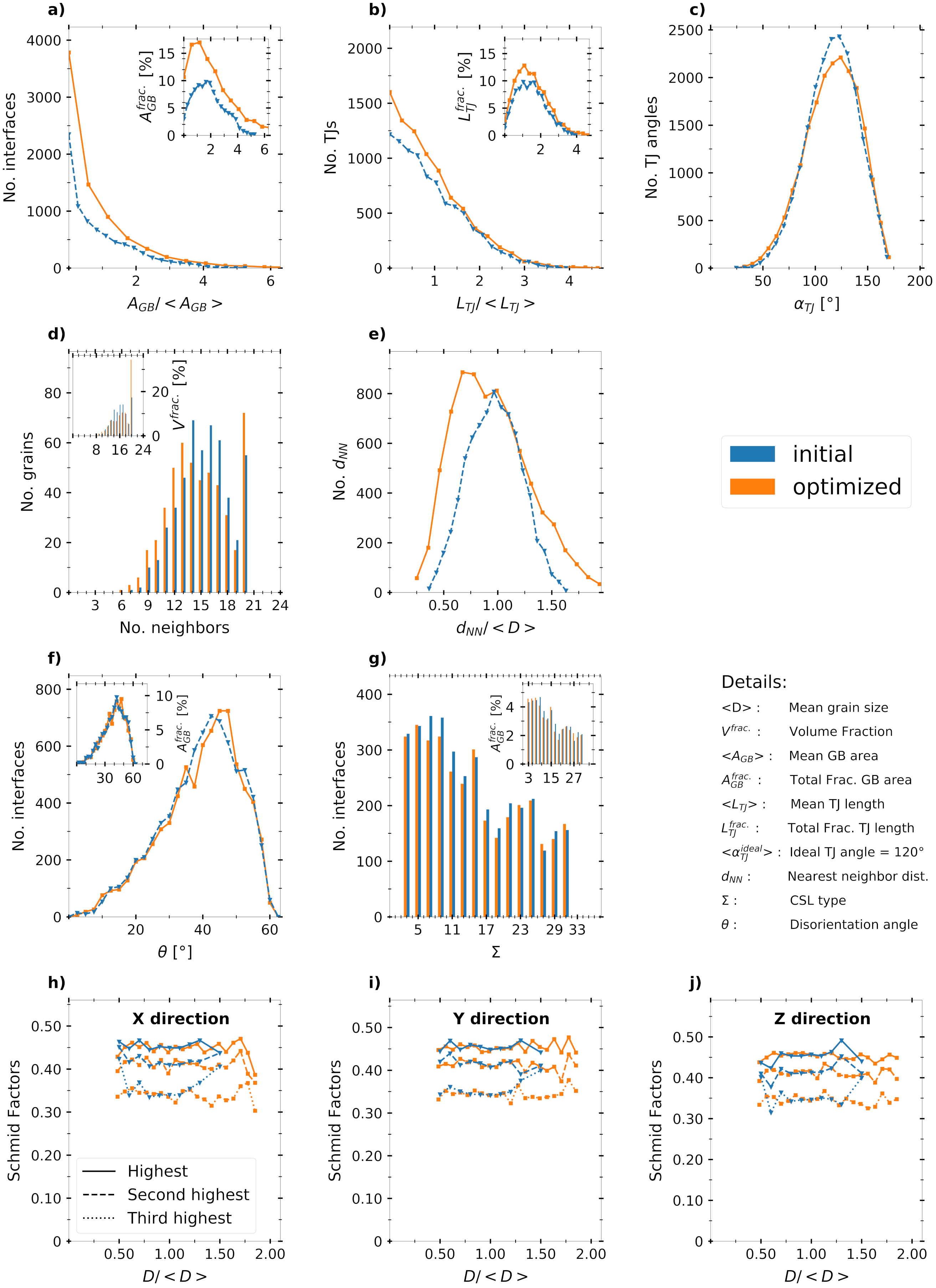} 
\caption{
Statistics of the 3D microstructure (example 1) shown as distributions: a) GB area; b) TJ length; c) TJ angle; d) Number of neighbors; e) Nearest neighbor distances; f) Disorientation between grains; g) Special GBs; h-i) First, second and third highest Schmid factors in x, y and z directions, respectively. For computing the Schmid factors, only the $\left[1 1 1 \right] \left<110\right>$ system is considered.
} 
\label{fig:3DmicroStats}
\end{figure}

Various characteristics of the microstructure generated by \software{} are shown in \fref{fig:3DmicroStats}. 
These statistics indicate that the optimization process results in an increase of small GBs and TJs (\fref{fig:3DmicroStats}a,b). The distribution of the angles at TJs (\fref{fig:3DmicroStats}c), however, remains almost identical. The optimization process also results in slight changes in the neighborhood with a mean of 12 neighbors observed in the optmized configuration. Additionally, the distribution of the nearest neighor distances is also broader, with a peak value around 75\% of the mean grain size.

No siginificant changes can be observed in the textural statistics. The disorientation distribution remains close to a Mackenzie type distribution \cite{mackenzie1957} after optimization, which is expected in a polycrystal with random orientations. A slight change of less than 1\% is observed in the number of $\Sigma 9, \Sigma 11, \Sigma 17, \Sigma 19$ GBs. The distribution of Schmid factors in individual grains, considering the global $x$, $y$ and $z$ as loading directions, also shows no significant changes.

\subsection{2D microstructure with user-defined cost function}\label{sec:2Ducf}
Generation of 2D-columnar microstructures with \software{} follows in principle the same procedure as in the 3D case. The seed coordinates are, however, defined only in the global $x$ and $y$ directions. The grains are finally extruded in the $z$ direction to obtain a columnar microstructure of a particular thickness.

\begin{figure}[htbp!]
\includegraphics[width=0.95\textwidth]{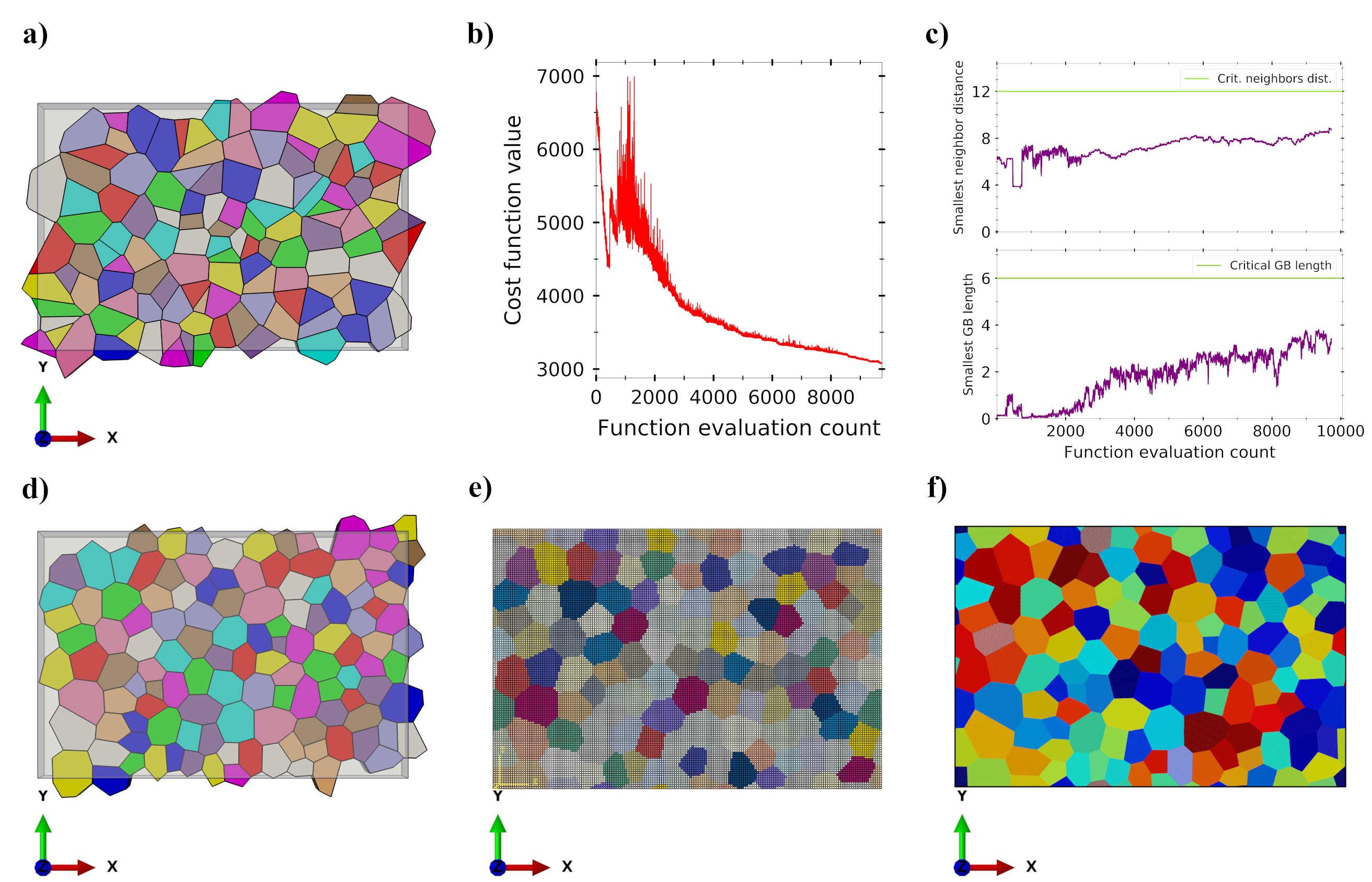} 

\caption{Generation of a 2D microstructure in a box of \mbox{180 x 120 x 15} units using a custom cost function: a) Initial microstructure with random seeds; b) Evolution of the overall cost during optimization; c) Evolution of the minimum neighbor distance and minimum GB area during optimization; d) Final optimized microstructure; e) Hexahedral FE mesh; f) Atomistic configuration with fcc crystal structure and $\left<110\right>$ orientation along the thickness. Color code as in \fref{fig:3Dmicro}.
} 
\label{fig:2Dmicro}
\end{figure}

Voronoi tessellated microstructures constructed with random grain centers can result in extremely small \mbox{GBs/TJs} (see \fref{fig:2Dmicro}a) which pose a problem during meshing of the configuration. A common method to overcome this problem is regularization, which involves the elimination of such \mbox{GBs/TJs} and collapsing corresponding vertices to a single vertex \cite{neper}. This, however, results in spurous topological features such as quintuple or higher order junctions, as against triple junctions that are generally observed in real microstructures, and can lead to erroneous results particularly in atomistic simulations \cite{hirvonen2017energetics}.

To overcome such undesired features, we define a cost function based on GB areas (here, lengths) and neighbor distances. The function is formally expressed as:

\begin{equation}
\mathcal{C} = f(\bar{d} - \textrm{d}^\textrm{c}) + f(\bar{A}_{GB} - \textrm{A}^\textrm{c}_{GB}),
\end{equation}
where $\bar{d}$ and $\bar{A}_{GB}$ denote the arrays of neighbor distances and GB areas, respectively. The parameters $\textrm{d}^\textrm{c}$ and $\textrm{A}^\textrm{c}_{GB}$ denote critical values for distance and GB area, and are taken to be 12 and 6 units, respectively. The function $f(\cdot)$ takes the form:
\begin{equation}
f(\cdot) = \sum_{1}^{n} \exp \left[ -\alpha * (\cdot) \right] + \beta * (\cdot), 
\end{equation}
where the summation is performed over the given array. The symbol $*$ indicates the product of two terms. The parameters $\alpha$ and $\beta$ work as scaling constants and are given values of 0.6 in the current work.

The exponential term results in repulsion of points that are very close to each other, whilst the linear term leads to an attraction of points that are far from one another, albeit to a lesser extent than the repulsion of close points. Upon optimization, the reduction in the overall cost function (\fref{fig:2Dmicro}b) results in an increase in the smallest nearest neighbor distance to approximately 8~units. Simultaneously, the smallest GB length increases tto approximately 2.5~units ((\fref{fig:2Dmicro}c). The optimized configuration shown in \fref{fig:2Dmicro}d exhibits no spurious features that are visible in the initial configuration (see supplementary movie S2 for more details). Each grain is then assigned a $\left<1 1 1\right>$ orientation along the thickness and a random in-plane rotation. A fcc crystal structure is used for generating individual crystals.

The command used for this example is:
\begin{lstlisting}[language=bash]
$> <@\highlight{./optimic main --size}@> 180 120 15 <@\highlight{--dimension}@> 2 <@\highlight{--number\_seed}@> 122 
   <@\highlight{--target}@> user_grain_size_distribution.txt <@\highlight{--characteristic}@> 2 
   <@\highlight{--characteristic}@> 5 <@\highlight{--material}@> example_2_2D_10k_so 
   <@\highlight{--stress\_direction}@> 1 0 0 <@\highlight{--sharp\_orientation}@> 1 1 1 
   <@\highlight{--seed\_spacing}@> random_3d <@\highlight{--rand\_seed}@> 2 
   <@\highlight{--optimization\_method}@> COBYLA <@\highlight{--max\_iter}@> 10000 <@\highlight{--number\_bins}@> 10 
   <@\highlight{--user\_cost\_func}@> user_cost_function_2.py 
   <@\highlight{--mesh}@> hex <@\highlight{--mesh\_size}@> 1.0
\end{lstlisting}

\begin{figure}[htbp!]. 
\includegraphics[width=0.95\textwidth]{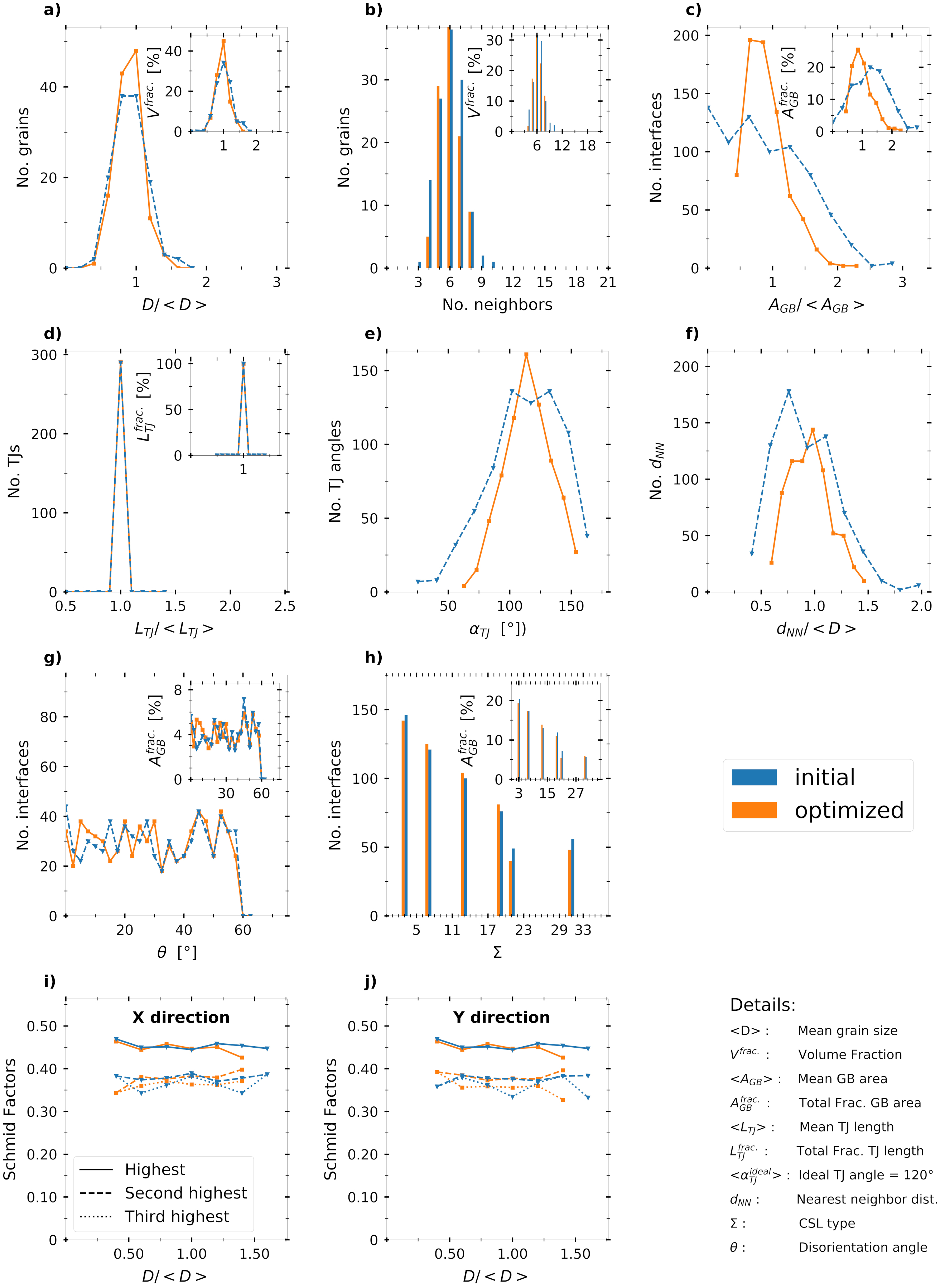} 
\caption{
Statistics of the 2D microstructure (example 2) as distributions: a) Grain size; b) Number of neighbors; c) GB areas; d) TJ lengths; e) TJ angles; f) Nearest neighbor distances; g) Disorientation; h) Special GBs; i,j) First, second and third highest Schmid factors of the $\left[1 1 1 \right] \left<110\right>$ family for loading in x and y directions.
} 
\label{fig:2DmicroStats}
\end{figure}

The statistics of the microstructure before and after optimization are shown in \fref{fig:2DmicroStats}. The optimization has almost no effect on the grain size distribution (cf. \fref{fig:2DmicroStats}a). The neighborhood distribution (\fref{fig:2DmicroStats}b) shows that grains with less than 3 and more than 8 neighbors are eliminated during the optimization process. Furthermore, very small and very large GBs are reduced. The distribution in TJ lengths shows no change since the length of any TJ is identical to the thickness of the film. The TJ angle distribution (\fref{fig:2DmicroStats}f) is more compact with reduced numbers of both large and small angles at TJs, indicating the manifestation of circular grains (cf. \fref{fig:2Dmicro}d) and is also seen in the distribution of neighbor distances (\fref{fig:2DmicroStats}f). No large differences can be observed in the textural statistics shown in \fref{fig:2DmicroStats}g-i.

\subsection{Optimized TJ angle distribution in a monodispersive microstructure}
\begin{figure}[htbp!]
\includegraphics[width=0.99\textwidth]{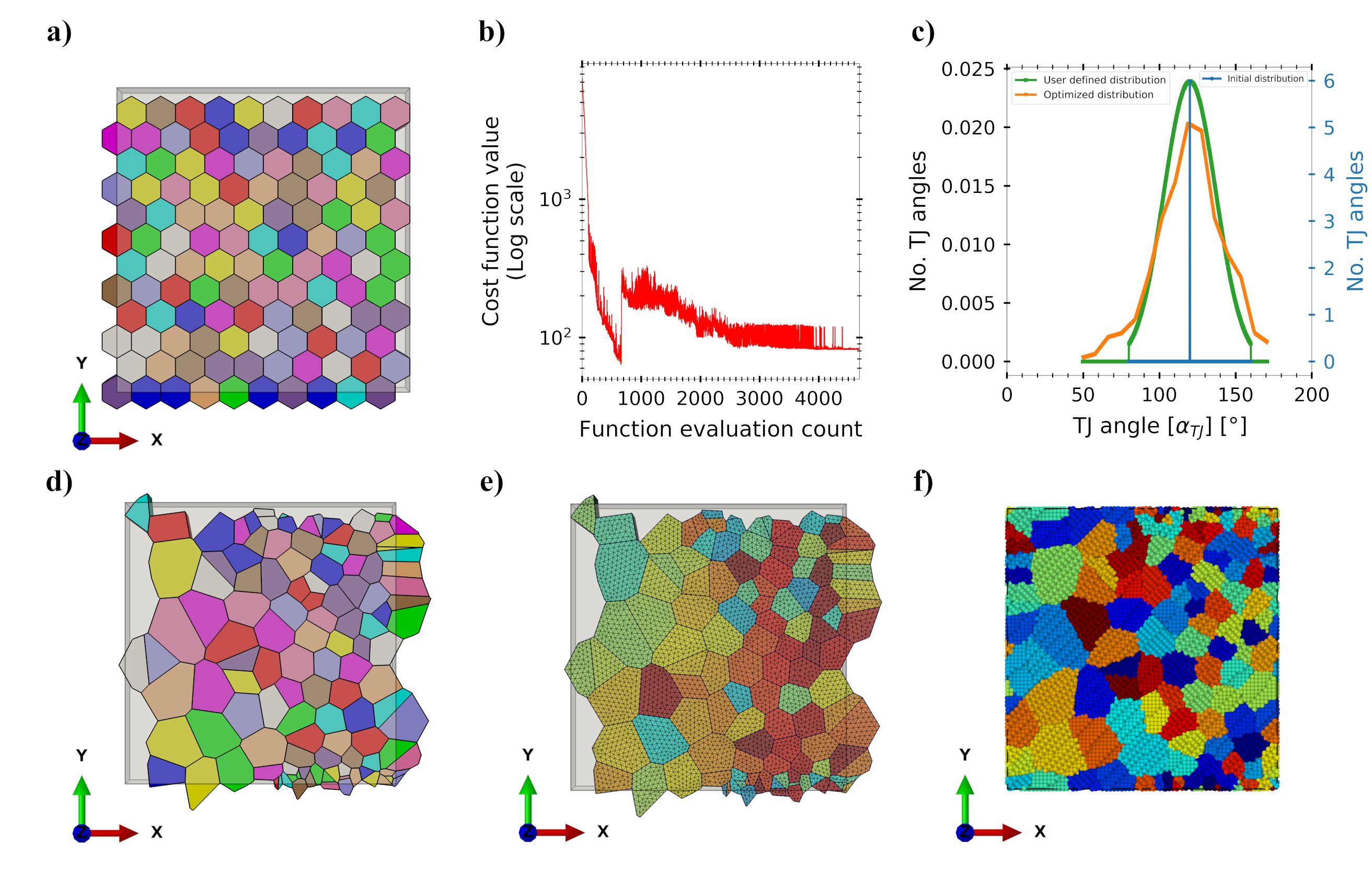} 

\caption{Optimization of TJ angle distribution in a mono-dispersive microstructure: a) Initial hexagonal microstructure obtained by seeds placed at regular intervals; b) Evolution of the cost function during optimization; c) Distribution of TJ angles; d) Optimized microstructure; e) Tetragonal FE mesh; f) Atomistic configuration with fcc crystal structure and $\left<110\right>$ orientation along the thickness. Color code as in \fref{fig:3Dmicro}.} 
\label{fig:2DmicroTJ}
\end{figure}

In this example, we demonstrate the optimization of a 2D microstructure starting from a mono-dispersive tessellation (hexagons) (\fref{fig:2DmicroTJ}a). The aim is to obtain a microstructure that looks more realistic than a simple hexagonal tessellation. To this end, we define a target distribution of TJ angles with a spread of $\pm 20^\circ$ around a mean TJ angle of $120^\circ$ found in the initial microstructure. The command used is:

\begin{lstlisting}[language=bash]
$> <@\highlight{./optimic main --size}@> 10. 10. 1. <@\highlight{--spacing\_length}@> 1 
   <@\highlight{--dimension}@> 2 <@\highlight{--target}@> junc_angle_80160_sc_17.txt 
   <@\highlight{--characteristic}@> 4 <@\highlight{--material}@> example_3_2D_hcp_longer 
   <@\highlight{--stress\_direction}@> 1 0 0 <@\highlight{--sharp\_orientation}@> 1 1 1 
   <@\highlight{--seed\_spacing}@> hcp_2d <@\highlight{--rand\_seed}@> 3 <@\highlight{--optimization\_method}@> COBYLA 
   <@\highlight{--max\_iter}@> 7000 <@\highlight{--number\_bins}@> 15 
   <@\highlight{--user\_cost\_func}@> user_cost_function_5_junc_angle.py <@\highlight{--mesh}@> tet 
   <@\highlight{--mesh\_size}@> 0.2
\end{lstlisting}

\begin{figure}[htbp!]
\includegraphics[width=0.99\textwidth]{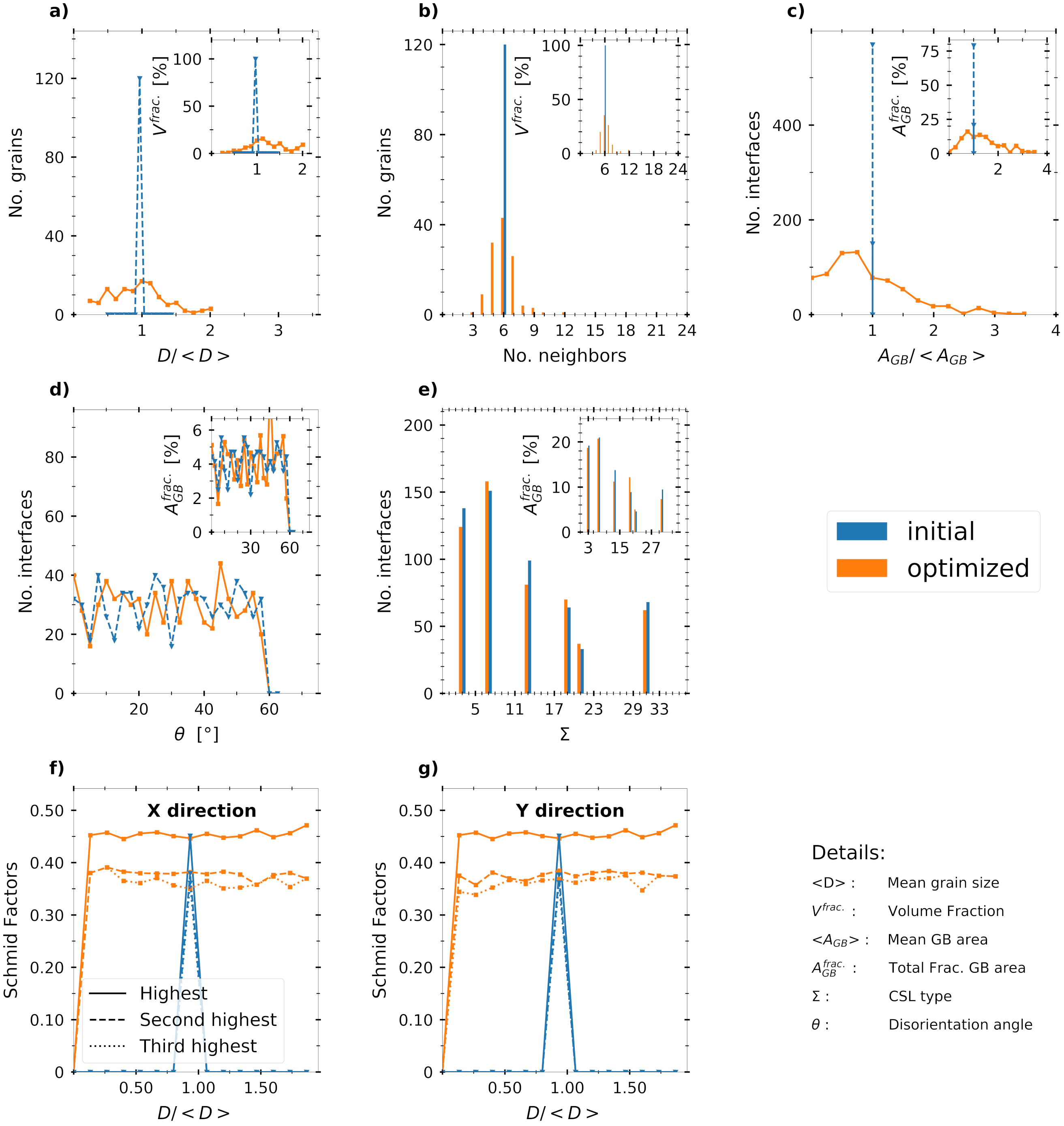} 

\caption{Statistics of the optimized 2D microstructure starting from a mono-dispersive tessellation shown as distribution: a) Grain size; b) Neighborhood; c) GB area; d) Disorientation angle; e) Special GBs; f,g) First, second and third highest Schmid factors of the $\left[1 1 1 \right] \left<110\right>$ family for loading in x and y directions, respectively.} 
\label{fig:2DmicroTJstats}
\end{figure}

The reduction of the cost function and the final distribution of the TJ angles are shown in \fref{fig:2DmicroTJ}b,c respectively, and as supplementary movie S3. The optimized Voronoi tessellation (\fref{fig:2DmicroTJ}d) shows grains of different sizes and shapes, unlike example 2 where the optimized tessellation manifested more-or-less circular grains. In this example, the meshed configuration is generated as a conforming mesh, i.e. mesh respecting the boundaries of individual grains, as seen in \fref{fig:2DmicroTJ}e. The atomistic configuration generated using a fcc crystal structure and $\left<110\right>$ orientation along the thickness is shown in \fref{fig:2DmicroTJ}f. For the sake of completeness, a few relevant statistics are shown in \fref{fig:2DmicroTJstats}.

%
%
%
%
%

\section{Conclusions}
In this work, we have developed an open source software \software{} that enables the generation of optimized microstructures for simulations of polycrystalline material behavior. The tool is able to generate both mono-dispersive as well as Voronoi microstructures, which can further be optimized to obtain certain key features and statistics seen in real microstructures. \software{} is highly flexible and allows users to define customized cost functions using a wide variety of statistics that are computed by the program. The functioning of the software has been demonstrated using three different examples, and encompassed both 3D and 2D microstructures, Voronoi and monodispersive microstructures, and the usability of custom cost functions. The code is programmed in a modular fashion and allows easy extension to other tessellation methods.

\section{Open Source Repository}
\textit{OptiMic} is available as open-source code and can be downloaded from the official \href{https://gitlab.com/arun.prakash.mimm/optimic}{\highlight{Gitlab }}\cite{optimicRepo} repository. The code is distributed under GNU General Public License, version 3.0.

\section{Conflict of Interest}
The authors confirm that there are no known conflicts of interest associated with this publication and there has been no significant financial support for this work that could have influenced its outcome.

\section{Acknowledgements}
SS acknowledges funding from the ERC starting grant, ``A Multiscale Dislocation Language for Data-Driven Materials Science'', ERC Grant agreement No. 759419 MuDiLingo.

\section*{References}


\end{document}